\documentclass[11pt,a4paper]{article}
\usepackage{amsmath,amssymb}
\usepackage{epsfig,graphicx}

\pdfoutput=1
\usepackage{eepic,overpic}
\usepackage{verbatim}

\let\oldbibliography\thebibliography
\renewcommand{\thebibliography}[1]{%
  \oldbibliography{#1}%
  \setlength{\itemsep}{0pt}%
}

\begin{document}

\title{\bf The Pierre Auger Observatory -- a new stage in the study of the ultra-high energy cosmic rays}
\author{Serguei Vorobiov$^a$\footnote{{\bf e-mail}: sergey.vorobyev@p-ng.si},
for the Pierre Auger Collaboration$^{b}$\footnote{{\bf e-mail}: info@auger.org.ar}
\\
$^a$ \small{\em Laboratory for astroparticle physics, University of
Nova Gorica} \\
\small{\em Vipavska 13, POB 301, SI-5001 Nova Gorica, Slovenia} \\
$^b$ \small{\em Pierre Auger Observatory} \\
\small{\em Av. San Mart\'in Norte 304, (5613) Malarg\"ue, Mendoza, Argentina}
}

\date{}
\maketitle

\begin{abstract}
This paper presents the highlights from the recent measurements
of the energy spectrum, mass composition, and arrival directions 
of the ultra-high energy ($\geq 10^{17}$~eV) cosmic rays (UHECR) 
by the Pierre Auger Observatory. The Auger hybrid detector combines 
fluorescence observations of the longitudinal profiles of extended air showers, 
initiated by these extremely energetic particles, with measures of the shower 
secondaries at the ground level by its large array of Cherenkov water tanks.
The complementariness of the two techniques provides important
cross-checks at the energies unreachable with accelerator experiments. 
The Pierre Auger experiment has observed a strong steepening of the
cosmic ray flux above $4 \times 10^{19}$~eV. A significant anisotropy 
in the arrival directions of cosmic rays has been established 
for events above $\simeq 6 \times 10^{19}$~eV. These events correlate 
over angular scales of less than $6^\circ$ with the directions towards
nearby ($D < 100$~Mpc) active galactic nuclei. Both observations
are consistent with the standard scenarios of the UHECR production
in the extra-galactic astrophysical acceleration sites. The derived 
stringent upper limits on the photon and neutrino contents 
in the UHECR are additional important arguments in support of these
scenarios. The measurements of the variation of the depth of
shower maximum with energy, interpreted with the current hadronic
interaction models, favour the mixed cosmic ray composition
at energies above $4 \times 10^{17}$~eV. We describe the
plans and prospects for the future, that will further help to solve 
the 50 years old UHECR puzzle.
\end{abstract}

\section*{Introduction}

After nearly a century of investigations, cosmic rays (CR) 
still represent a scientific challenge and offer 
fundamental questions about their origin and nature. 
The CR energy spectrum remarkably shows only little 
deviation from a constant power law $J(E) \propto E^{-2.7}$ 
across the large energy range from $10^{9}$~eV to $10^{20}$~eV. 
The main paradigm for explanation of this non-thermal 
spectrum is the diffusive acceleration of charged 
particles in the vicinity of astrophysical shocks. 
The small variations of the spectral index are signs
of possible changes in the CR acceleration and/or 
propagation regimes. The significant increase in the slope 
of the CR spectrum to $E^{-3.0}$ occurs at the so-called knee 
near $5 \times 10^{15}$~eV, and is presently attributed 
to a maximum energy expected in the diffusive shock 
acceleration in Galactic supernova remnants (SNR).  
At much higher energies, corresponding to the ``ankle'' 
observed at $\sim 4 \times 10^{18}$~eV, the spectrum 
becomes again $E^{-2.7}$. The integral cosmic ray flux 
above the ankle is only about 1~particle per km$^2$ steradian year,
which implies the need for huge detectors to collect 
substantial statistics. 

Supernova remnants are thought to be prime candidates for the acceleration 
of galactic cosmic rays up to energies of several $10^{17}$ eV.
At these ultra-high (UHE, E $\geq 10^{17}$~eV) energies, 
only a few astrophysical  acceleration sites
unify the conditions (sufficiently large size, strong magnetic 
fields) allowing to efficiently produce the cosmic rays,
a list of candidates being limited to objects like active 
galactic nuclei, gamma-ray bursts or galaxy clusters.
Coincidentally, the cosmic rays above $ \geq 10^{17}$~eV
~are able to arrive on Earth from regions external to the Local Group of
galaxies (D $\sim 3$~Mpc).
Above $ \geq 10^{19}$~eV, the bulk of the UHECR
is believed to be extragalactic, which is comforted
by the non-observation of anisotropies connected with the Galactic
plane. 

If this standard acceleration scenario with the nuclei arriving 
from sources at cosmological distances holds, it is expected 
that the cosmic ray flux should undergo a strong suppression 
(so-called GZK cut-off~\cite{gzk66}) at energies 
above~$\sim 5 \times 10^{19}$~eV. This spectral break 
corresponds to the effective threshold of pion production 
in the interaction of the UHECR protons with the CMB radiation. 
At similar energies, nuclei photo-dissociate on the CMB. 
As a consequence, the horizon of the UHECR sources in the standard 
scenarios is restricted to our local ``neighbourhood'' ($\simeq$ 50 Mpc).      
Above the GZK cut-off, the deflections of cosmic rays
in the galactic and intergalactic magnetic fields
are minimal, and individual CR sources should be observed. 

Alternatively, more exotic UHECR scenarios have been proposed.
In some of these scenarios, cosmic rays
are produced in interactions or decays of the primordial Universe 
relics such as topological defects or super-heavy dark matter. 
Contrary to the astrophysical acceleration scenarios, 
these ``top-down'' UHECR production models predict a substantial 
fraction of primary photons and neutrinos.

To distinguish between different UHECR scenarios, it is crucial 
to measure precisely the cosmic ray arrival direction distribution, 
primary mass composition, and energy spectrum. 
Since the UHECR fluxes are small, these highest energy 
particles are only detectable in an indirect way, 
through the extensive air showers (EAS) they create in the atmosphere.
Above the energies around $10^{14}$~eV/nucleon, 
the technique of sampling the shower particles, predominantly electrons, gammas, 
and muons, by detectors on the ground becomes efficient. 
Lateral distributions of EAS particles, measured by surface detector arrays,
reflect the particle densities in the tail of the shower 
far away from the shower axis, and serve to estimate 
the primary energy and mass. However, at the highest energies such estimates 
depend strongly on assumptions about the hadronic interactions,
which are extrapolated to energies exceeding by far 
those accessible at man-made accelerators and in regions of phase space 
not covered in collider experiments.
The advantages of surface detector arrays include stable operation with 100\%
percent duty cycle, and a relatively straightforward determination of
the aperture.

In the UHE regime, where the Auger experiment operates, 
the air fluorescence technique allows estimation of the primary 
CR energy. As fluorescence light is produced proportional to the
energy deposited in the atmosphere, this technique provides 
a calorimetric energy measurement that is fairly independent 
of the details of the shower development, and thus of our 
understanding of the individual hadronic interactions. 
Unfortunately, the fluorescence technique can only be used in dark, 
clear nights, which reduces its duty cycle to about 10\%.
Another drawback of this technique is that the aperture
of a fluorescence detector grows with shower energy, and its
determination for all operating conditions is nontrivial.

During the long period before the advent of the Pierre Auger Observatory, 
the world largest cosmic ray data set above $10^{19}$ eV has been collected 
by two experiments, each exploiting only one of these
methods, either the detection of shower particles on the ground 
(AGASA surface array), or the fluorescence technique (HiRes). Consequently, 
the results of AGASA and HiRes were affected by different 
systematics. In addition, because of the small available statistics,
the key issue of the presence of the GZK cut-off in the UHECR energy 
spectrum remained unclear.
To improve this situation, the Auger Observatory 
was conceived in the 1990s~\cite{AugerDesignReport} as a hybrid detector,
combining the two detection techniques and covering 
an area 30 times larger than that of the AGASA array.

\section*{The Pierre Auger Observatory}

\begin{figure}[!ht]
\centering
\includegraphics[width=0.6\textwidth]{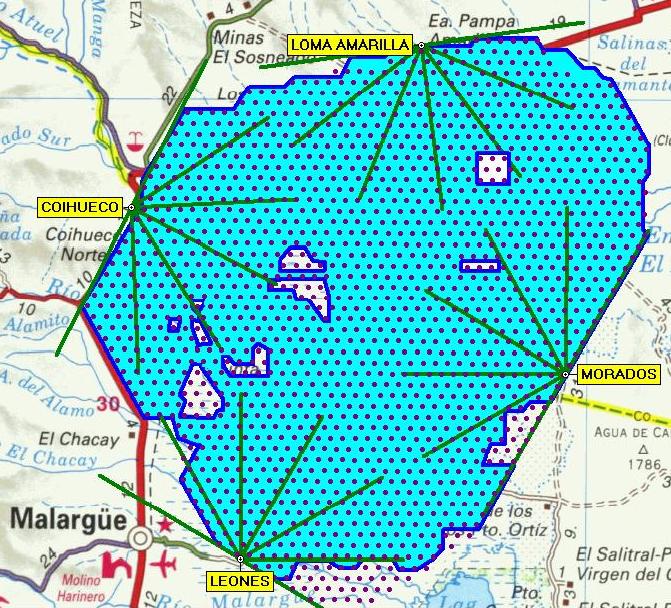}
\caption{Deployment status of the Pierre Auger Observatory, as of June 2008.
The points on the hexagonal grid show the positions of the
water Cherenkov detectors. The shaded area marks the surface detectors in
operation. The four named fluorescence detectors with six individual 
telescopes at the perimeter of the surface array are all operational.
The central campus is located in the town of Malarg\"ue.
}
\label{fig:sd-array-status-20080611}
\end{figure}

The Pierre Auger Observatory for the ultra-high energy
cosmic ray studies has been designed and built by an 
international collaboration of more than 350 scientists 
representing $\sim 100$ institutions from 17 countries.  
The Observatory site in the province of Mendoza, Argentina 
will be completed in the year 2008. The Auger Surface Detector (SD) consists of 1600 
water Cherenkov detectors, deployed over 3000 km$^{2}$ on a hexagonal 
grid with 1500 m spacing. The SD array is overlooked by the Auger
Fluorescence Detector (FD), which comprises 24 fluorescence
telescopes, arranged in four sites at the perimeter of the SD area.
This hybrid design allows the simultaneous detection of the same 
cosmic ray events by two complementary techniques, which provides 
important cross-checks and measurement redundancy.
The Auger Detector layout and the status of deployment as of June 2008 
is shown on the Fig.\ref{fig:sd-array-status-20080611}. 
The properties and performance of the hybrid instrument 
can be found in~\cite{augernim,AugerHybridPerformance}.

Each Surface Detector station consists of a polyethylene tank, 
1.55 m high and 3.6 m diameter, enclosing a liner filled with 12 tons 
of purified water. Three 9 inch diameter photomultiplier tubes (PMTs),
located symmetrically at a distance of 1.2 m from the center of the tank,
look downward through windows in the liner and collect the Cherenkov 
light from the passage of showers particles in the water. 
The PMT signals are digitized at~$40~\rm{MHz}$ sampling frequency, 
which provides a temporal resolution of~$8~\rm{ns}$. 
The electronics package also includes GPS receiver and 
a radio communication system. Each tank is equipped with a 10 Watt 
solar power system and operates in an autonomous
manner. A robust, automatic calibration procedure of the SD
detectors \cite{AugerSDCalibration} uses the measurement 
of the average charge, collected 
by a PMT from the Cherenkov light produced by relativistic muons 
passing vertically through the center of a tank.
Subsequent measures of the signal are given in such vertical 
equivalent muon (VEM) units. The SD calibration procedure
is carried out continuously, and enables to determine 
the signals recorded from extensive air showers with 3\% accuracy. 
It also allows to achieve stable and uniform triggering conditions
through the whole SD array. 

A hierarchical trigger sequence is applied to ensure the rejection of 
accidental triggers in order to retain the physical events.
Then the tank signal timing information and the integrated charge 
values are used for reconstruction of shower geometry (arrival 
direction and core position). The measured signals as a
function of distance from the shower axis are fitted for individual
events to obtain the water Cherenkov signal at a distance of 1000 m, 
S(1000), which serves as an estimator of the size of each event.
Simulations show that both for proton- and iron-induced showers  
the S(1000) parameter is almost proportional to primary energy. 
Also, at these distances from the axis the fluctuations of lateral 
distribution of particle density are minimal for studied energies 
and adopted array geometry, for zenith angles $\theta < 60^\circ$.
The overall uncertainty in S(1000) at energies around $10^{19}$ eV 
is about 10\%, and improves with energy \cite{AugerSDReconstructionAccuracy}. 
For S(1000) $< 40$ VEM (which corresponds to energies 
lower than~$\sim 6 \times 10^{18}$ eV),
the main contribution to the uncertainty is due to the finite
dimension of the detectors and the small counting statistics. 
At higher energies, the major contribution to the uncertainty 
in S(1000) arises from the fluctuations in the shower development 
and the consequent lack of knowledge of the true lateral particle 
distribution shape for a particular event.

The angular resolution of the surface detector was determined
from the shower front arrival times, on an event by event basis, 
and checked using the data from the pairs of SD stations located 
$\sim 11$~m apart (see \cite{AugerSDReconstructionAccuracy} and
the references therein).  The resolution was found to be
better than 2$^{\circ}$ for 3-fold events (events with 3 stations
participating in the event, which corresponds to
energies $E < 4 \times 10^{18}$ eV), better than 1.2$^{\circ}$ 
for 4-fold and 5-fold events ($3 \times 10^{18} < E < 10^{19}$ eV) 
and better than 0.9$^{\circ}$ for higher multiplicity events 
($E > 10^{19}$ eV).

The ``quality'' cuts, which are used in most analyses, 
select among the physical events those with the highest signal 
recorded in a tank surrounded by six operating ones, 
and with the reconstructed core within a triangle of working stations.
Also, the events with arrival zenith angles $\theta < 60^\circ$
are considered. These selection criteria guarantee that showers 
are sufficiently well contained within the SD array to allow 
a robust measurement of S(1000) and the shower axis. The quality cuts 
result in a 100\% combined trigger and reconstruction efficiency 
above $3 \times 10^{18}$ eV \cite{AugerSDExposure}. At higher energies, 
the effective area of the SD array at any time is calculated 
from the number of active hexagons, which can be determined from 
the low-level triggers sent by each station every second
(see \cite{AugerSDExposure} and the references therein).\\

The Auger Fluorescence Detectors measure the faint ultra-violet 
light emitted as the shower traverses the atmosphere.
Each of the 24 individual Schmidt design telescopes 
covers a~$30^\circ$ range in azimuth and~$0^\circ \div 30^\circ$ 
range in elevation. A~$11~\rm{m}^2$ segmented spherical mirror
(radius of curvature~3.4~$\rm{m}$) is focusing the light from
the~$2.2~\rm{m}$ diameter diaphragm onto a camera of~$20\times22$
PMTs, each with a field of view of 1.5$^\circ$ in diameter.
The PMT signals are digitized at~$100~\rm{MHz}$ sampling
rate. At the diafragm opening, there is a ring corrector lens
to double the light collection efficiency and to reduce
the aberrations, and a UV-transparent filter window, which also
serves as a protection from dust.   

The optical and electronics calibration of the FD system
is known to 9.5\%. The absolute end-to-end calibration is performed
periodically through the year, using a uniformly illuminated
drum positioned at the diafragm aperture. The NIST calibrated
light source with known intensity and spectral and directional 
characteristics allows for measurements at five wavelengths. 
The relative calibration, made twice every night with pulsed 
LEDs and/or Xe flashers, keeps track of any changes between the drum 
calibrations.

The atmosphere is an integrated part of the FD operation,
and the attenuation of the fluorescence light due to Rayleigh 
and aerosol scattering along its path toward the telescope
have to be carefully monitored.
Each FD site is equipped with a single wavelength LIDAR. 
Regular balloon launches are performed to measure vertical
profiles of the air temperature and pressure.
The atmospheric monitoring system also includes one Raman
LIDAR, a meteo station per FD site, a remote-controlled
centrally positioned YAG laser, and 
dedicated aerosol monitors and cloud cameras. 
The systematic uncertainties in atmospheric attenuation contribute
approximately 4\% to the systematic uncertainty for
shower energy estimate. For the energy deposited in the
atmosphere by shower particles the data from \cite{NaganoEtAl2004}
are currently being applied, with the systematic uncertainty 
in the absolute fluorescence yield of 14\%. The measurements 
from \cite{Airfly2007} are used for the wavelength and pressure 
dependence of the fluorescence spectrum. Uncertainty related
to pressure, temperature and humidity effects amounts to 7\%.

The image of a shower developing in the field of view of a
telescope represents a track of triggered PMTs, which enables to
reconstruct a shower-detector plane (SDP) with a high precision ($\simeq
0.3^\circ$). The pixel timing information is used to determine the
shower direction. When, in addition to an FD telescope, one or more SD 
tanks participate in the event (hybrid detection), the SD timing
information improves considerably the shower geometry reconstruction.
Compared to the case of mono FD events, the accuracy of angular 
reconstruction and of the determination of the core position 
of the showers are both improved about a factor of 10.

In the algorithm used to reconstruct the longitudinal profile 
of a shower, care must be taken to collect the fluorescence light
without inclusion of the night-sky background, and to properly
account for a contribution of direct and scattered Cherenkov light.  
The reconstructed profile of the energy deposit, fitted 
with a Gaisser-Hillas function \cite{Gaisser-Hillas}, provides a measurement 
of the shower energy, and of X$_\mathrm{max}$, 
the depth of the shower maximum. Systematic uncertainties in 
the reconstruction method contribute 10\% to the total uncertainty in
the measured energy. A final small correction ($\simeq 10\%\pm4\%$ at 
the Auger energies) takes into account the ``missing'' energy 
due to muons and neutrinos.

The quality cuts applied to the hybrid events selected for 
the data analysis are described in \cite{AugerHybridComposition}.
The hybrid method provides a nearly calorimetric, model-independent 
energy measurement with a statistical resolution of 8\%. 
The corresponding resolution in X$_\mathrm{max}$ is 20 g/cm$^{2}$.
The quoted systematic uncertainties in the energy scale by 
the hybrid method (see \cite{AugerHybridPerformance} and the references
therein) add up to 22\%, the largest contributions being due to the
uncertainties in the absolute fluorescence yield (14\%) and the
absolute FD calibration (10\%).\\

The results reported here are essentially based on the analysis of the
data collected at the Pierre Auger Observatory between 1 January 2004 
and 31 August 2007. During this time, the SD array grew from 154 to 
1388 tanks, and the number of FD telescopes increased from 6 to 24.  
The integrated exposure over this period is four times greater than 
that of AGASA, similar to the monocular HiRes exposure, and is 
equivalent to roughly 1 year of operation of the full Auger array.
Above $10^{18}$ eV, the Auger Observatory has recorded more events 
than all previous experiments together. The expected future annual 
increment of the exposure is 7000 km$^2$ sr yr.

\section*{The UHECR energy spectrum}

\begin{figure}[!t]
\centering
\begin{overpic}[width=0.48\textwidth]{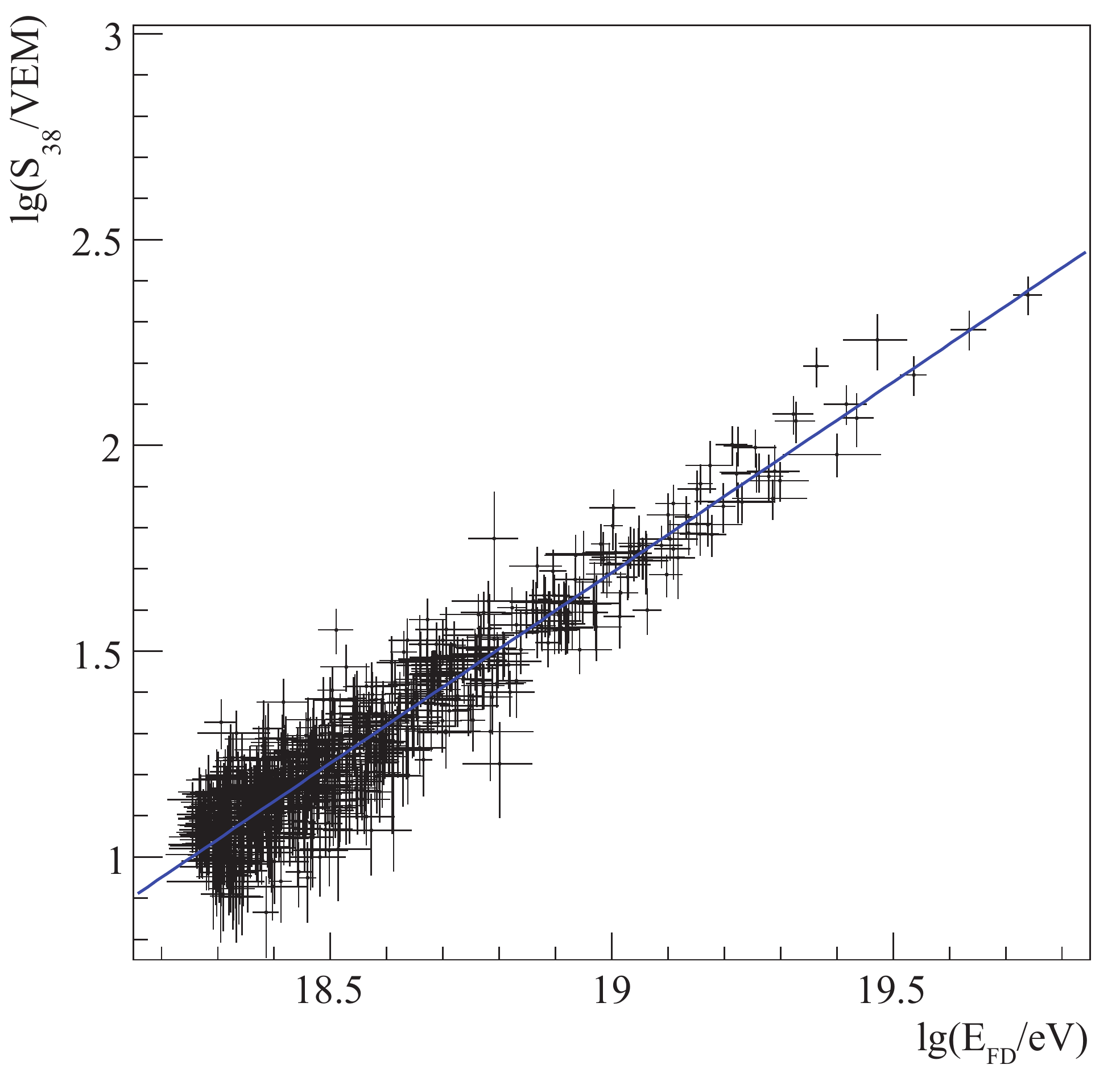}
\put(15.8,52.5){\includegraphics[width=0.205\textwidth,clip=]{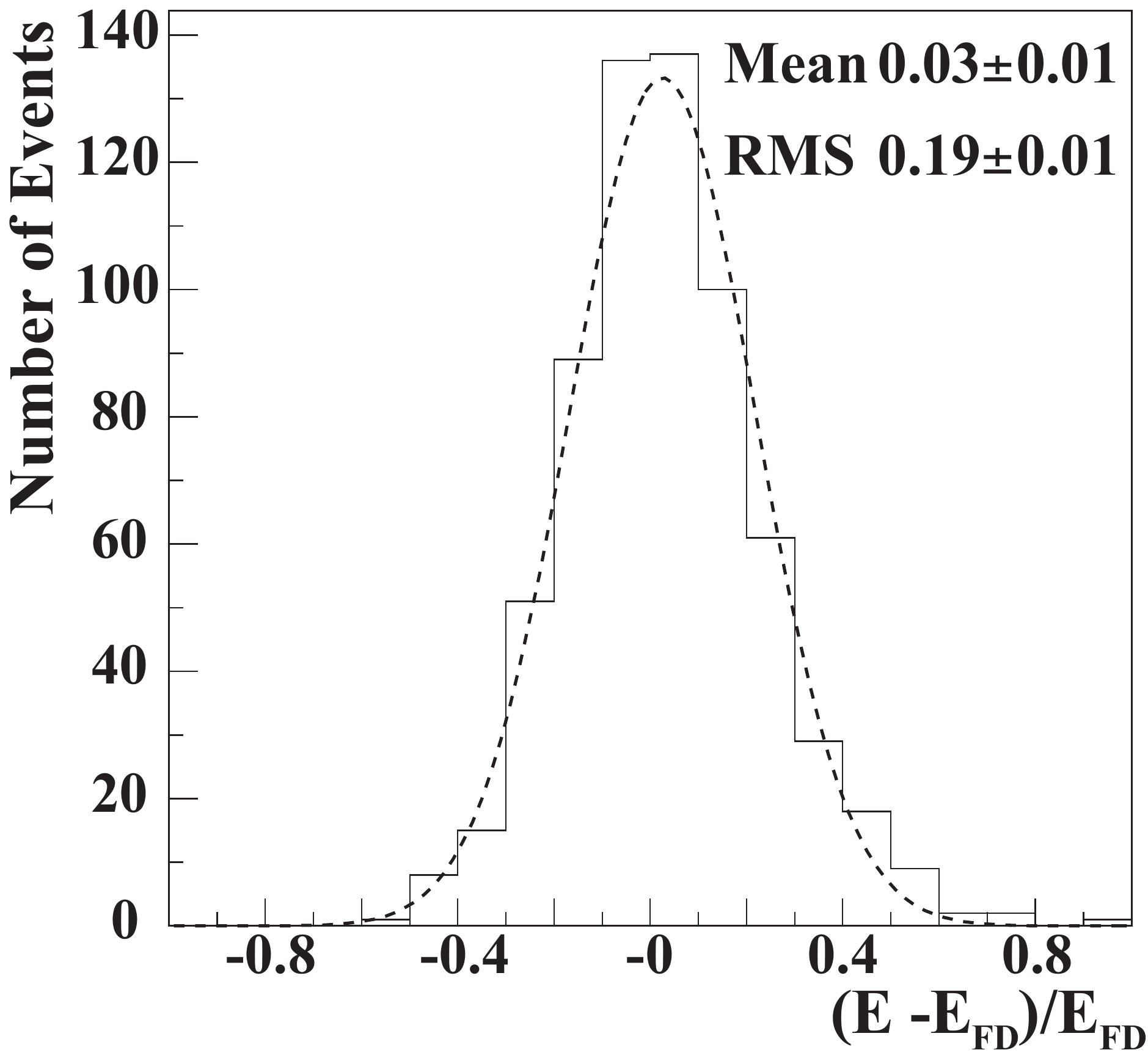}}
\end{overpic}
\includegraphics[width=0.48\textwidth]
{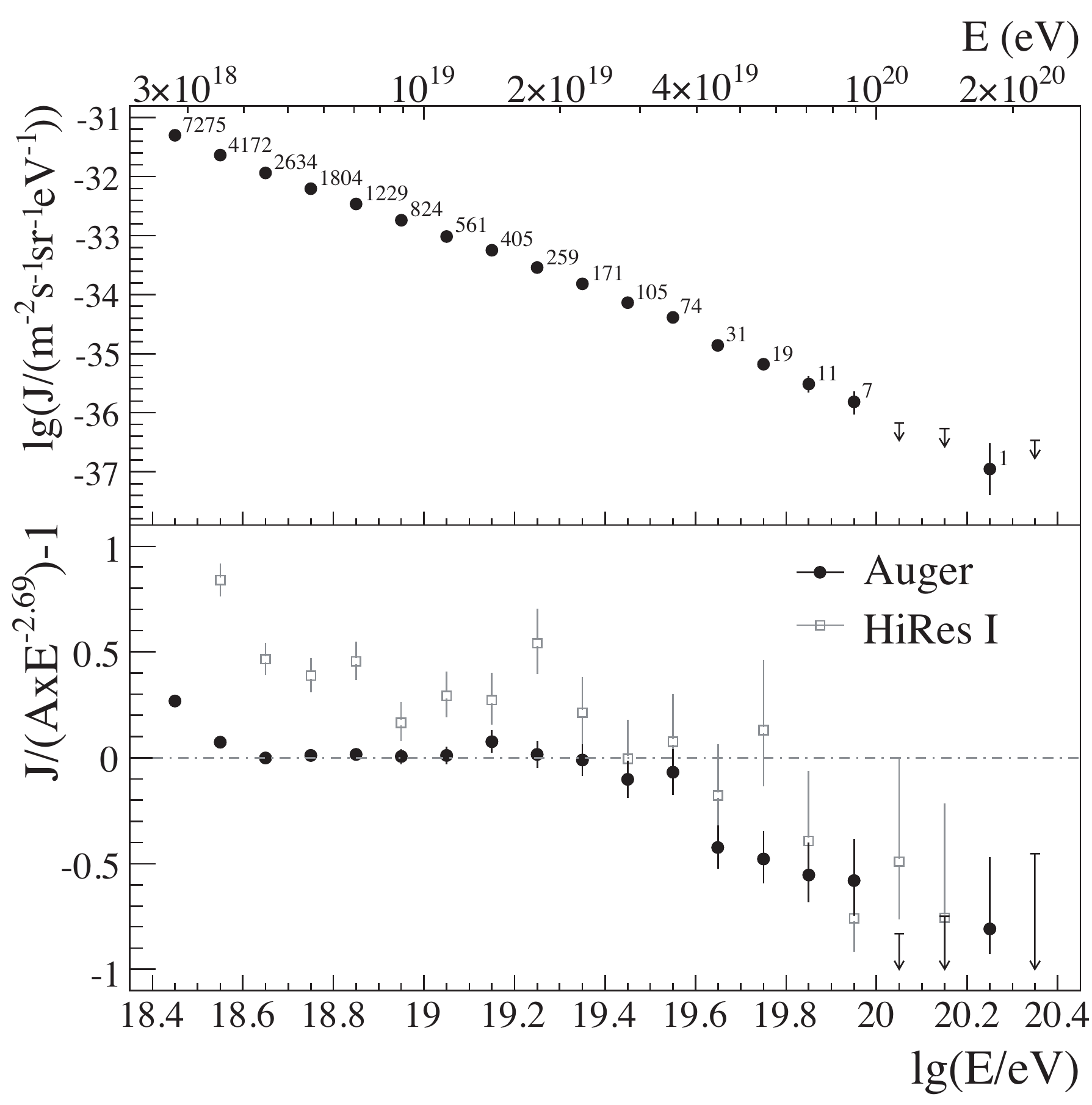}
\caption{The energy calibration by the hybrid method and the resulting
Auger UHECR energy spectrum \cite{AugerSpectrumPRL}:
\newline \textit{(left)} 
Correlation between SD energy parameter $S_{38^{\circ}}$ and FD energy
for the 661 hybrid events used in the fit. The full line is the best
fit to the data; the fractional differences between the two energy 
estimators are shown in the inset.
\newline \textit{(right)} 
Upper panel: Differential flux as a function of energy, with 
statistical uncertainty. The number of events in each bin is also
shown. 
Lower panel: Fractional differences between Auger and HiRes I 
data compared to a spectrum with an index of 2.69.
\label{fig:s38-fdenergy-correlation-energy-spectrum}}
\end{figure}

The hybrid nature of the Auger detector allowed a measurement
of the energy spectrum to be made with high statistics, 
and with small dependence on assumptions about 
hadronic interaction models or the primary mass composition.
We use the constant integral intensity cut method, which exploits 
the nearly isotropy of cosmic rays, to rescale S(1000) value 
from different zenith angles $\theta$. The choice of the threshold 
S(1000) value is not critical, since the shape of the attenuation
curve giving intensity I($>$S(1000)) in equal solid angle 
bins of $\cos^2 \theta$ is within the statistics nearly the same 
in the large range of S(1000) \cite{AugerSDSpectrum2007}.
We normalize S(1000) for each event to the signal it would have 
produced at a zenith angle of 38$^{\circ}$, which is the median value 
for zenith angles of interest. We establish the relation between
that quantity called $S_{38^{\circ}}$, and the calorimetric FD energy 
measurement for currently 661 selected high-quality hybrid events.  
A clear correlation between the two energy estimators can be seen 
on the left plot in 
Fig. \ref{fig:s38-fdenergy-correlation-energy-spectrum}.
A power law fit to the data is performed, yielding 
$E_\mathrm{FD}=1.49\times 10^{17} \mathrm{eV}\times S_{38^{\circ}}^{1.08}$,
which shows that $S_{38^{\circ}}$ grows approximately linearly with
energy. The energy resolution, estimated from the r.m.s.
deviation of the distribution, is 19\%, which is in good agreement 
with the quadratic sum of the SD and FD energy statistical
uncertainties of 18\%.

The energy spectrum based on 20,000 SD events is shown on 
the right plot in Fig. \ref{fig:s38-fdenergy-correlation-energy-spectrum}.  
The systematic uncertainty in the absolute energy scale set by the FD is 22\%, 
as described in the previous section. The spectrum between
$4\times10^{18}$ eV and $4\times10^{19}$ eV is well approximated
by a power law with the slope $-2.69\pm0.02(\mathrm{stat})\pm0.06(\mathrm{syst})$,
where the systematic uncertainty comes from the calibration curve.
If this slope is extrapolated to higher energies, the expected
number of events above $4\times10^{19}$ eV and $10^{20}$ eV
is $167\pm3$ and $35\pm1$, while 69 events and 1 event are 
observed. The Auger data therefore clearly show that the slope
of the spectrum increases above $4\times10^{19}$ eV, with 
the significance of the steepening being more than 6 standard
deviations. Above this energy, the spectral index is 
$-4.2\pm0.4(\mathrm{stat})\pm0.06(\mathrm{syst})$.
On Fig. \ref{fig:s38-fdenergy-correlation-energy-spectrum}, 
fractional differences relative to a spectrum $J(E) \propto E^{-2.69}$
are also shown, together with the data from the HiRes I 
experiment \cite{HiResSpectrum2008}. The AGASA data are not shown,
since they are currently under revision \cite{teshimaRICAP2007}.

The Auger UHECR energy spectrum measurements 
have been extended to the zenith angle range between $60^\circ$ and $80^\circ$.
A dedicated reconstruction procedure has been implemented 
\cite{AugerHorizontalReconstruction} for such inclined 
events, dominated by muons. The shower axis
and the total number of muons are reconstructed 
using simulation-based maps of the muon ground density.
The hybrid Auger events are used in the similar way as above
to establish the relationship between the muon number 
and the energy measured by FD. A total of 734 events above 
$6.3\times10^{18}$ eV were used to build the spectrum 
\cite{AugerHorizontalSpectrum}. The integrated exposure 
for this measurement is 1510 km$^2$ sr yr, which corresponds 
to 29\% of the respective exposure for events below $60^\circ$.

The Auger spectrum has also been extended to lower energies
down to $10^{18}$ eV, using hybrid events \cite{AugerHybridSpectrum}.
The three energy spectrum measurements (using SD vertical, SD inclined
and hybrid events) are consistent where they overlap. 
The astrophysical implications of the measured
UHECR energy spectrum that features 
flux suppression above $4\times10^{19}$ eV and the steepening 
below the ankle at $\sim 4\times10^{18}$ eV (see 
Fig. \ref{fig:s38-fdenergy-correlation-energy-spectrum})
are discussed in \cite{AugerSpectrumImplications}.

\section*{Mass composition studies using X$_\mathrm{max}$}

\begin{figure}[!ht]
\centering
\includegraphics[width=0.48\textwidth]{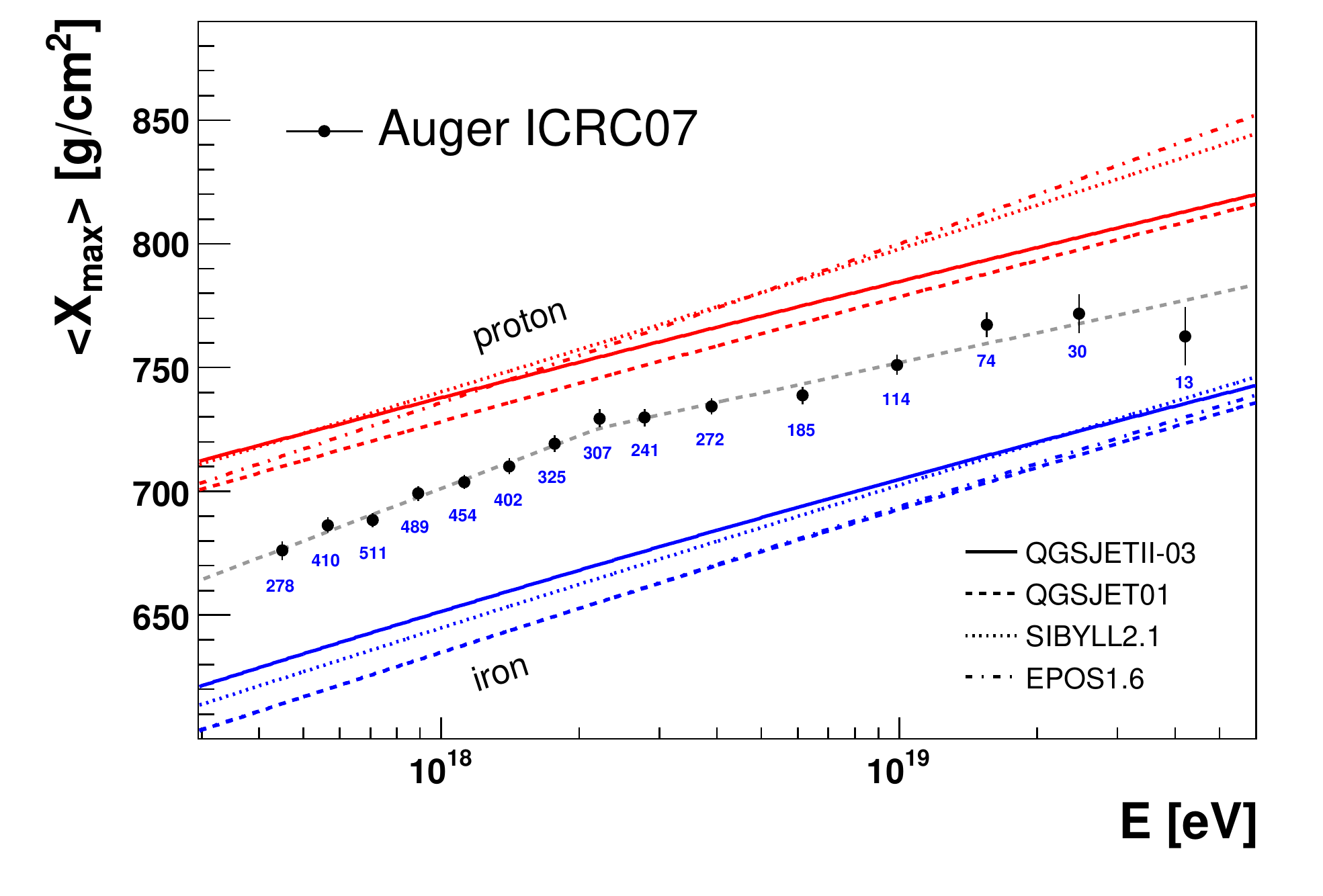}
\includegraphics[width=0.48\textwidth]{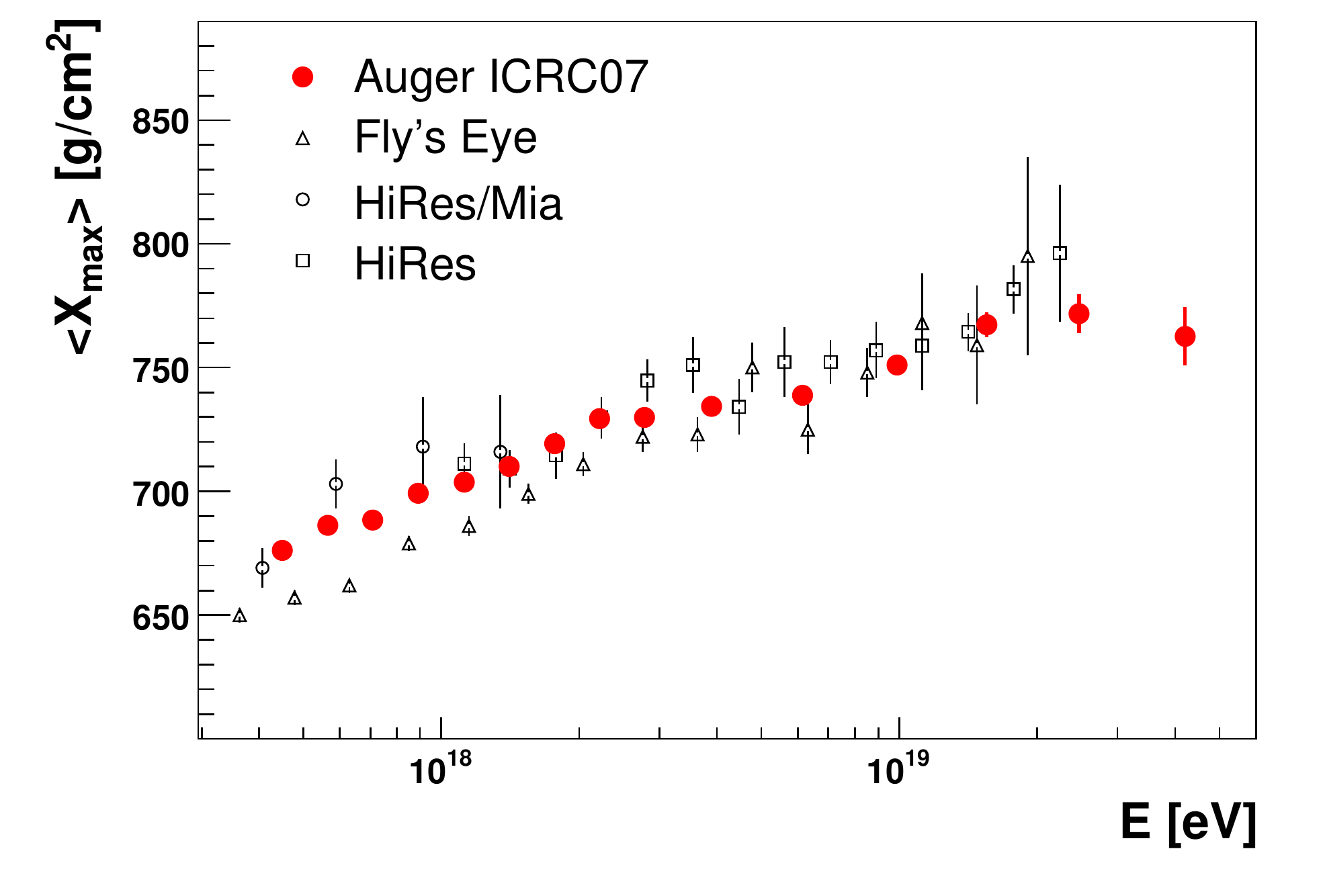}
\caption{
Auger measurements \cite{AugerHybridComposition} of the average depth 
$\langle \mathrm{X}_\mathrm{max} \rangle$ of shower maximum 
as a function of energy.
\textit{(left)}
$\langle \mathrm{X}_\mathrm{max} \rangle$ compared to predictions 
from hadronic interaction models for primary protons and iron 
nuclei. The dashed line shows results of a fit with two constant 
elongation rates and a break-point. 
The number of events in each bin is indicated below the data points.
\textit{(right)}
$\langle \mathrm{X}_\mathrm{max} \rangle$ compared to forerunner experiments.
}
\label{fig:xmax-energy-models-experiments}
\end{figure}

The hybrid technique allows for precise measurement of 
the depth of shower maximum as a function of energy.
Heavy nuclei are expected to reach the maximum development at smaller
average depths than protons and to produce 
smaller shower-to-shower fluctuations.
The cosmic ray mass composition can therefore be studied
by comparing the observed average X$_\mathrm{max}$ 
with predictions from air shower simulations for different nuclei. 
The change of average X$_\mathrm{max}$ with energy 
is used to probe the changes in primary composition.  
The Auger result \cite{AugerHybridComposition} 
based on 4329 hybrid events across two decades of energy
is shown on Fig. \ref{fig:xmax-energy-models-experiments}.
The left plot compares the Auger measurements with
predictions for proton and iron primaries made
using three models of hadronic interactions.
If the interaction models are correct, the Auger data
favour mixed composition for all energies above $4 \times 10^{17}$
eV. However, there are indications that the hadronic 
models may need modification \cite{AugerHadronicModels}.
There is a general agreement with the results from previous experiments, 
as it can be seen from the right plot on the same figure.
At the same time, the Auger measurement is more precise
and covers larger energy range. The apparent change 
from lighter to heavier mass composition above 
$2\times10^{18}$ eV is currently examined by studying 
the fluctuations in X$_\mathrm{max}$ at a given energy. 

\section*{The photon and neutrino limits}

The presence of photons and neutrinos in the UHECR is guaranteed due 
to the interactions of cosmic ray nuclei with background radiation
fields via the production of neutral and charged pions 
$\pi^0, \pi^\pm$ and their subsequent decays. A discovery of these
``GZK'' photons and neutrinos would provide a unique opportunity
to trace back their production sites, and to test fundamental particle
physics at energies well beyond current or planned accelerators.

\begin{figure}[!ht]
\centering
\includegraphics[width=0.48\textwidth]{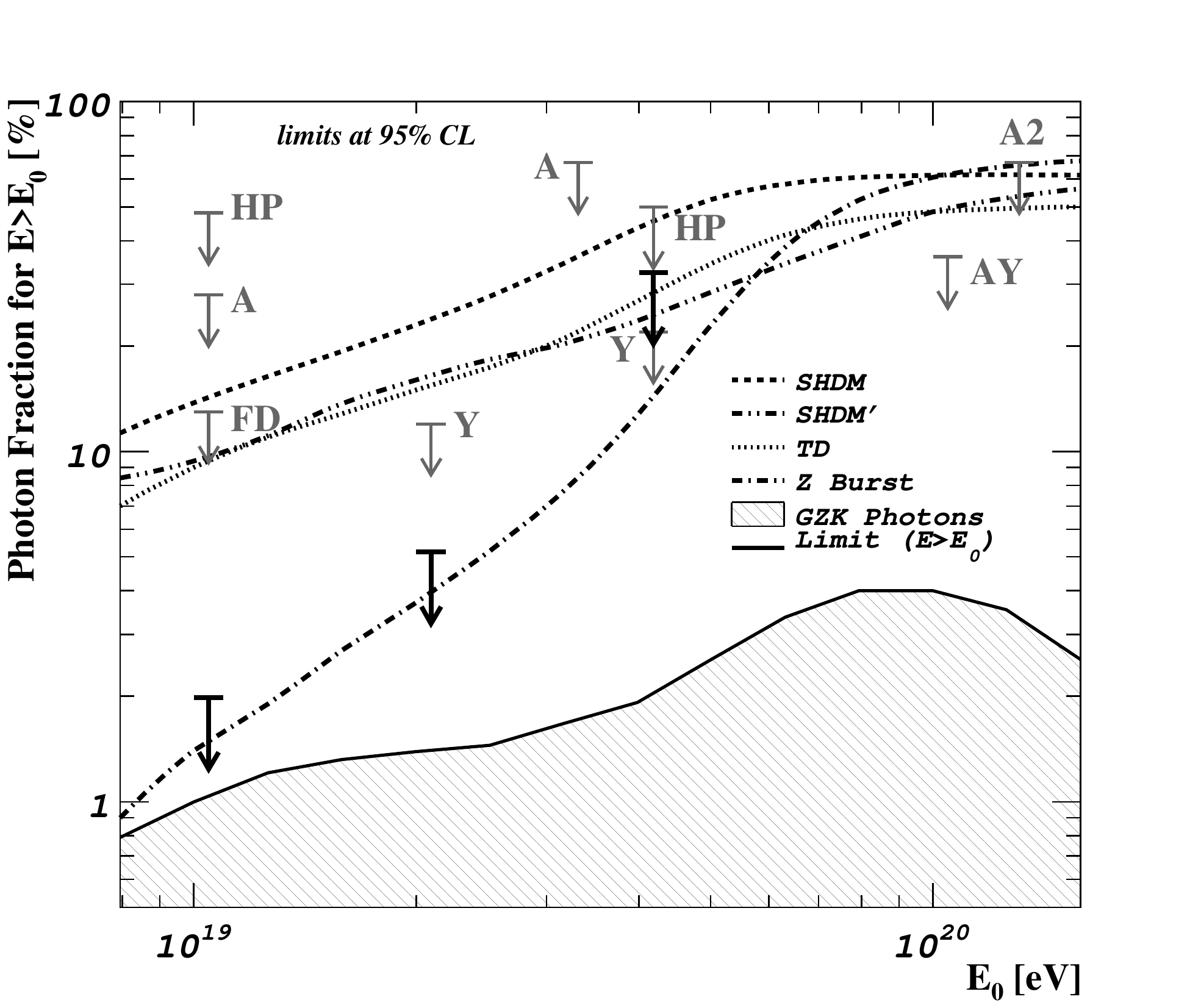}
\includegraphics[width=0.48\textwidth]{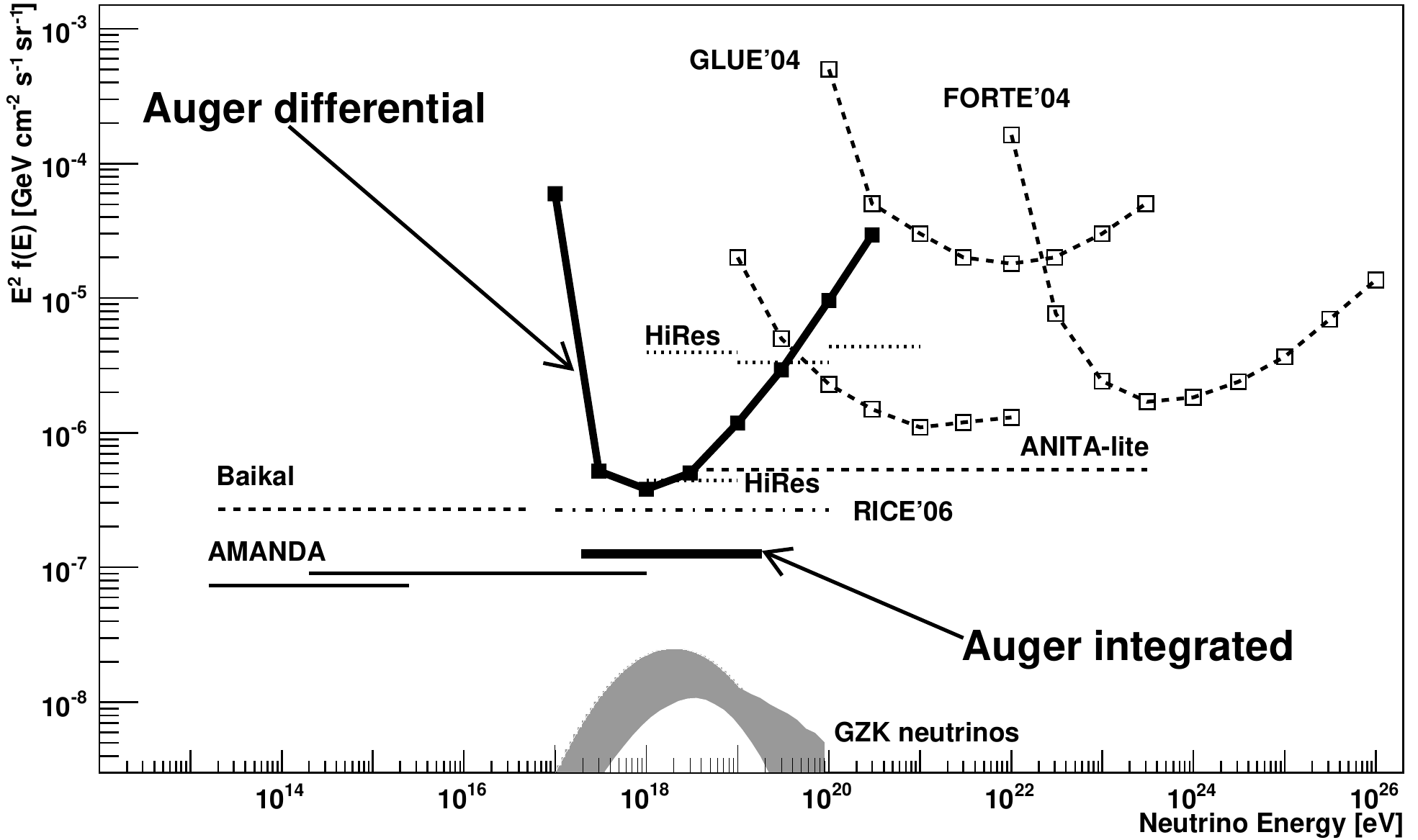}
\caption{Limits on UHE photon and neutrino contents.\newline
\textit{(left)}
The upper limits on the fraction of photons in the integral cosmic ray
flux derived from Auger SD events (black arrows) along with the previous
experimental limits (HP: Haverah Park; A, A1, A2: AGASA; AY:
AGASA-Yakutsk; Y: Yakutsk; FD: Auger hybrid limit). 
Also shown are predictions from top-down models and for the GZK photon
fraction. See \cite{AugerSDPhotonLimit} and the references therein.
\newline \textit{(right)}
Limits at 90\% C.L. for a diffuse flux of $\nu_\tau$ from the 
Pierre Auger Observatory. Limits from other experiments are converted 
to a single flavour assuming a 1:1:1 ratio of the 3 neutrino flavours 
and scaled to 90\% C.L. where needed. The shaded curve shows a typical
range of expected fluxes of GZK neutrinos, although predictions almost
1 order of magnitude lower and higher exist.  
The references can be found in \cite{AugerNeutrinoLimitPRL}, where
the figure is taken from.}
\label{fig:photon-fraction-neutrino-limits}
\end{figure}

Showers induced by the UHE photons develop deeper
in the atmosphere than those initiated by the primary nuclei, and
contain fewer muons. The first Auger limit on photon contents
in the UHECR had been derived exploiting the X$_\mathrm{max}$
measurement in hybrid events \cite{AugerFDPhotonLimit}. 
A more stringent limit has now been obtained \cite{AugerSDPhotonLimit}
using the large SD event statistics. Two photon-sensitive SD
observables have been combined in the analysis: the risetime of 
the recorded shower signal and the radius of curvature of the shower front.
Since photons are less efficient in triggering the Auger Surface
Detector, the analysis has been limited to energies higher 
than $10^{19}$~eV, and to zenith angles $\theta$ between
 $30^{\circ}$ and $60^{\circ}$. The photon energy scale has been 
estimated with a help of a custom reconstruction that accounted 
for the universality of the development of the electromagnetic
component of showers for depths exceeding X$_\textrm{max}$. 
A search for photon candidates in the Auger SD data has been
performed, and no candidate has been found. This allowed us 
to obtain upper limits on photon flux and fraction 
above 10, 20, and 40 EeV (1 EeV $\equiv 10^{18}$~eV) 
at 95\% confidence level (the respective 
limits on the photon fraction are 2.0\%, 5.1\%, and 31\%; 
the corresponding limits on the photon flux are (in units
of km$^{-2}$ sr$^{-1}$ yr$^{-1}$) 
$3.8 \times 10^{-3}, 2.5 \times 10^{-3}, \textrm{and} ~2.2 \times 10^{-3}$).
The resulting Auger upper limits (see the left plot 
on Fig. \ref{fig:photon-fraction-neutrino-limits}) represent
significant improvement upon the results from the previous
experiments. They strongly constrain exotic models of UHECR production.
With the accumulation of Auger data in the incoming years, 
the level expected for GZK photons may be reached.\\

A large air mass above the Auger SD array represents an important target
for detection of showers induced by UHE neutrinos. Such showers
can be identified if they develop deep in the atmosphere under large
zenith angles, by the presence of a significant electromagnetic shower component. 
Two types of neutrino events can be detected by the SD array: 
down-going showers produced by a flux of neutrinos 
of all flavours \cite{DowngoingNeutrinos}, 
and upcoming showers initiated by the decay of $\tau$s emerging 
from the Earth after the propagation of a $\nu_\tau$ 
flux \cite{SkimmingNeutrinos}. 
Identification criteria have been developed to find EAS 
due to the UHE Earth-skimming neutrinos. No candidates have been found
in the data collected between 1 January 2004 and 31 August 2007.
Assuming an E$^{-2}_{\nu}$ differential energy spectrum,  
we derive a 90\% upper limit on the diffuse tau neutrino flux as $E^{2}_{\nu} dN_{\nu_{\tau}} /
dE_{\nu} < 1.3 \times 10^{-7}$ GeV cm$^{-2}$ s$^{-1}$ sr$^{-1}$ in the
energy range $2 \times 10^{17}$ eV $< E_\nu < 2 \times 10^{19}$ eV.
Our result \cite{AugerNeutrinoLimitPRL}, which is at present 
the most sensitive bound on neutrinos in the EeV energy range, 
is shown on Fig. \ref{fig:photon-fraction-neutrino-limits}. 

\section*{Results on the UHECR anisotropies}

The search for anisotropies in the arrival directions
of cosmic rays has been a long-standing goal. 
If the observed suppression of the UHECR energy spectrum is due
to the cosmic ray interactions with the CMB, the CR arrival
directions at the energies above the suppression are expected to correlate 
with the nearby matter distribution, which is quite 
inhomogeneous. Observation of such correlations can constitute
a first step towards doing the cosmic ray astronomy, provided that 
cosmic ray deflections in the intervening galactic and 
extragalactic magnetic fields are small enough, so that
the CR arrival directions point back to their origin.

To test for possible correlations with extragalactic 
objects, the Pierre Auger Collaboration analysed
 \cite{AugerAGNCorrScience,AugerAGNCorrAPh}
the arrival directions of the events above $4\times 10^{19}$~eV, 
to search for coincidences with the positions of the
nearby ($D < 100$ Mpc) active galactic nuclei (AGN)
listed in the V\'eron-Cetty and V\'eron catalogue \cite{vcv}.
A scan over the maximum angular separation $\psi$ between 
the events and the AGNs, the maximum considered AGN 
redshift $z_\mathrm{max}$, and the lower threshold energy
$E_\mathrm{th}$ for cosmic ray events was performed to search 
for the most significant correlation. The range of the scanned 
parameter values has been chosen large enough,
given the unknown intervening magnetic fields and cosmic
ray mass composition (which would affect the amplitude 
of deflections, and also the GZK horizon distance) 
as well as the systematic uncertainties in energy determination.

A deep minimum in the probability $P$ of observing a similar
or larger number of correlations arising from simulated isotropic data
sets was observed for $\psi=3.2^\circ$, $z_{max}=0.017$ (or maximum AGN
distance of $71$~Mpc) and $E_{th}=57$~EeV (which corresponds to
the 27 highest energy events). Under these parameters, only $\sim 10^{-5}$ of  
isotropic simulations have a deeper minimum. A correlation was first
found in the data taken before the end of May 2006, for very
similar parameter values. A test with this set of parameters, fixed \textit{a priori}, 
has been applied to the subsequent data collected up to the end of
August 2007, and the anisotropy has been confirmed using this
independent data set at more than 99\% confidence level.

\begin{figure}[!ht]
\centering
\includegraphics[width=0.8\textwidth]{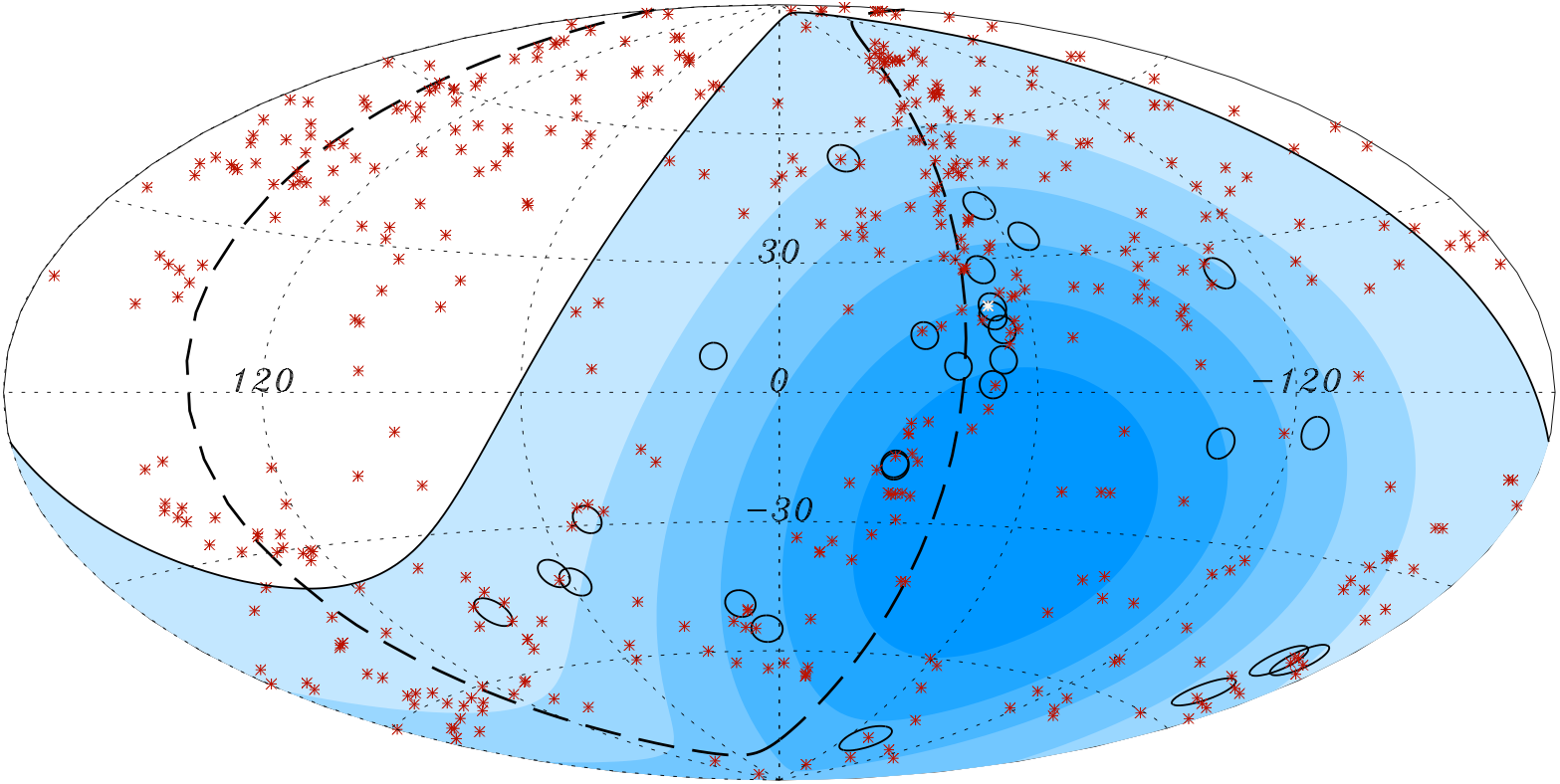}
\caption{Map in galactic coordinates
with circles of 3.2$^\circ$ radius centered on the arrival directions 
of the 27 Auger events with energy $E>57$~EeV (taken from 
\cite{AugerAGNCorrScience,AugerAGNCorrAPh}).
The positions of the 442 AGN from the 12$^{th}$ edition of the V\'eron-Cetty and 
V\'eron catalogue with redshifts $z \leq 0.017$ ($D < 71$~Mpc) are
shown with red stars (one of the closest objects, Cen A, is shown in white). 
Shading within the solid line (the Auger field of view for
$\theta<60^{\circ}$) indicates regions of equal exposure. 
The dashed line denotes the supergalactic plane.}
\label{fig:skymap-agn-correlation}
\end{figure}

The map of the arrival directions and of the AGN positions is shown on
Fig.~\ref{fig:skymap-agn-correlation}. Out of the 27 most energetic
events, 20 are at less than $3.2^\circ$ from an AGN closer than
71~Mpc, while only 5.6 events were expected on average if the CR flux were
isotropic. A remarkable alignment of several events with the
supergalactic plane is observed. Also, two events fall within 
3.2$^\circ$ from Centaurus~A, one of the closest active galaxies.
It is worth to note that the energy maximizing the correlation with
AGNs coincides with the one that maximizes the autocorrelation of the events
\cite{AugerClusteringStudies}, and also with the energy
for which the cosmic ray flux decreases to 50\% of the power law extrapolation 
(see Fig.~\ref{fig:s38-fdenergy-correlation-energy-spectrum}).

The observed correlation suggests that the highest energy 
cosmic rays originate from nearby extragalactic sources,
either AGN or other objects with a similar spatial distribution.
This result is consistent with the hypothesis that the measured 
steepening of the spectrum above 40 EeV is due to the GZK effect, 
rather than to a maximum energy attainable in cosmic ray
accelerators. The larger statistics and further investigations 
will help to unambiguously identify the UHECR sources.\\

At lower energies, the Pierre Auger Collaboration has performed search for 
extended and point-like sources in the Galactic Center region, 
and no significant excess has been found in the Auger data
\cite{AugerGCStudies}. Also, searches for large-scale
anisotropies have been performed by looking for patterns 
in the right ascension modulation of the cosmic ray distribution 
\cite{AugerLSAnisotropyStudies}. No anisotropy of this kind has been found, and
an upper limit on the first harmonic modulation of 1.4\% in the energy
range 1 $<$ E $<$ 3 EeV was set.

\section*{Conclusions and future prospects}
The recent Pierre Auger Observatory measurements of the UHECR energy spectrum, 
arrival directions and mass composition, albeit taken during 
the detector construction phase, have already significantly contributed 
to the progress in the field. The obtained results have demonstrated the power 
of the Auger hybrid design. The Southern Auger Observatory is now almost 
complete (the Surface Detector assembly and deployment activities 
have been essentially completed in August 2008). The Auger South 
will provide the good quality experimental data for at
least ten years.\\

Since the elaboration of the Auger Project design 
report~\cite{AugerDesignReport} in the 1990s, it was proposed to
construct two similar instruments in both hemispheres, in order to
cover the whole sky. In 2005, a site in Colorado, USA, has been
chosen to host the Northern Auger Observatory. The Auger North
will focus on the highest energy cosmic rays, with the aim to 
provide higher statistics at energies above the observed cosmic ray 
flux suppression. This will facilitate identification of individual 
nearby UHECR sources against the isotropic event background from 
the more distant objects, and thus will establish the long-sought 
charged particle astronomy.

To facilitate the data integration from both Observatories, 
Auger North will be built using the same basic
elements as in the Southern Observatory: water Cherenkov detectors, 
fluorescence telescopes, and an associated infrastructure 
for communications and calibration. The performed research 
and development work has led to the design~\cite{AugerNorth} 
of an array of $4,000$ surface detectors, covering the area of 
$8,000$ square miles, or $20,740 \, \mathrm{km}^2$, i.e. 
$\sim 7$ times the Southern array. From 40 to 50 fluorescence 
telescopes will be needed for the almost full hybrid coverage 
of this large area. The envisaged layout of surface 
detectors (positioned with the spacing of $\sim 2.3 \, \mathrm{km}$ 
on a square grid) will allow the array to be fully efficient for 
hadronic showers above $3 \times 10^{19}$~eV. The proposal for Auger
North is currently being finalized.\\ 

While it is clear that the highest energy cosmic rays 
arrive from extragalactic sources, it is currently not known 
at which energy the extragalactic CR flux becomes dominant. 
This question is central to the understanding of the nature of 
accelerators and their injection spectrum, as well as of the CR 
propagation in large-scale cosmic magnetic fields. Competing 
scenarios could be distinguished by better determination of CR mass 
composition and energy spectrum in the energy range 
between about $10^{17}$~eV and $3 \times 10^{18}$~eV,
where the transition between Galactic and extragalactic cosmic rays
is expected~\cite{AugerEnhancementsMotivation}.

Two hybrid ``low'' energy extensions of the Southern Auger 
Observatory (Auger Enhancements) will explore this energy range. 
New fluorescence detector, a system of three High Elevation 
Auger Telescopes (HEAT), is currently being installed near 
the Coihueco fluorescence site. The first telescope is expected 
to be operational in November 2008, and the last two 
will be operational in early 2009. HEAT will cover a range 
of elevation angles from 30$^\circ$ to 60$^\circ$. The combined 
shower data of the Coihueco and the HEAT telescopes will lead 
to better resolutions for the determination of air shower energy 
and X$_\mathrm{max}$, especially below $\sim 10^{18}$~eV~\cite{HEAT}.

To extend the hybrid technique down to $\sim 10^{17}$~eV,
a denser array of water Cherenkov detectors of present design is currently
being deployed on a $23.5 \, \mathrm{km}^2$ area centered 6 km 
away from the fluorescence detector installations at Coihueco site.
Each water Cherenkov detector will be accompanied by a 
$30 \, \mathrm{m}^2$ muon scintillator counter buried alongside
3 m underground. The whole complex, baptized AMIGA 
(Auger Muons and Infill for the Ground Array) will consist of 
85 pairs of water Cherenkov and muon detectors, 
placed on 433 and 750 m triangular grids.
AMIGA surface detectors will complement the fluorescence measurements,
and the muon detectors will be used to determine the mass composition 
in the range $10^{17} \div 5 \times 10^{18}$~eV~\cite{AMIGA}.\\

Within the Auger Collaboration, there is also an active program
to investigate the radio emission of extensive air showers
in the 10 $\div$ 100 MHz band~\cite{Radio}. This method for shower studies, 
first introduced in the 1960s, has revived recently due to 
the new radio measurement techniques. 
The radio shower detection is exploitable with a 100\% duty cycle,
and provides additional information on the development 
of the electromagnetic shower component, 
which can be used to measure the primary CR energy,
arrival direction, and mass in a way complementary to the
SD and FD techniques. The tests of radio detection of showers 
with various antenna systems are currently being performed 
on the Southern Auger site. A recently obtained funding will allow 
the construction of a $20 \, \mathrm{km}^2$ engineering array
at the location of AMIGA infill. This engineering array 
will serve to the development of radio detection, with a goal
to elaborate a design of a much larger array, that will cover
an area of many thousand $\mathrm{km}^2$.

\section*{Acknowledgements}

I would like to thank the organizers of the 15$^{\mathrm{th}}$
International Seminar "QUARKS-2008", and of the accompanying
4$^{\mathrm{th}}$ International UHECR Workshop, for inviting me 
to present the results of the Pierre Auger Collaboration 
at both conferences. 
The support from the University of 
Nova Gorica is gratefully acknowledged.

\end{document}